\documentclass[twocolumn,showpacs]{revtex4}
\pdfoutput=1
\usepackage{amsmath}
\usepackage{amssymb}
\usepackage{graphicx}
\usepackage{subfigure}
\usepackage{nicefrac}
\usepackage{pdfsync}

\newcommand{\abs}[1]{\left|#1\right|}

\begin{document}
\title{Three Balls Problem Revisited -- On the Limitations of Event-Driven Modeling}

\author{Patric M\"uller}
\author{Thorsten P\"oschel}
\affiliation{  Institute for Multiscale Simulation,
  Universit\"at Erlangen-N\"urnberg,
  N\"agelsbachstra{\ss}e 49b,
  Erlangen,
  Germany }
\date{\today}
\begin{abstract}
If a tennis ball is held above a basket ball with their centers
vertically aligned, and the balls are released to collide with the floor, the tennis
ball may rebound at a surprisingly high speed. We show in this article
that the simple textbook explanation of this effect is an
oversimplification, even for the limit of perfectly elastic particles. Instead,
there may occur a rather complex scenario
including multiple collisions which may lead to a very different final velocity as compared with the 
velocity resulting from the oversimplified model.
\end{abstract}

\pacs{45.70.-n,45.70.Qj,47.20.-k}

\maketitle

\section{Introduction}

\begin{figure}
  \centering
  \includegraphics[width=0.6\columnwidth]{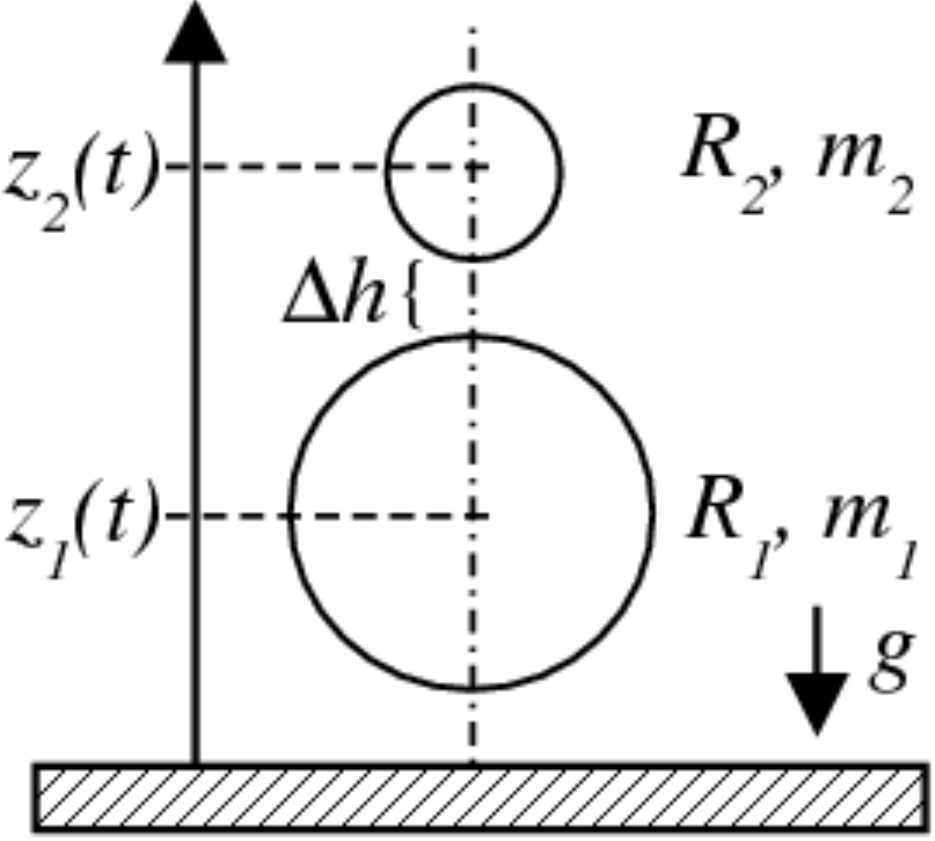}
\caption{Sketch of the problem.}
  \label{fig:sketch}
\end{figure}

Consider a set of two balls made of the same viscoelastic material whose centers
are vertically aligned at positions $z_1$ and $z_2$ as sketched in Fig.
\ref{fig:sketch}.
Let $R_1$ and $R_2$ be the radii of the particles and $\Delta h$ their initial
vertical spacing. At time $t=0$ we release the
particles to collide with the floor. The question for the maximum
height reached by the upper sphere after the collision is then a
common textbook problem. The textbook solution is based on the
assumption that the collisions of the lower particle with the floor
and the subsequent collision of the lower particle with the upper one
are separate two-body interactions which may be treated independently,
that is, one disregards the duration of the collisions. Whether this
assumption is justified depends, of course, on the initial conditions,
the particle sizes and the material parameters. 

The experimental investigation of the described problem is tricky as
even a small deviation from the vertical alignment of the initial
positions of the spheres leads to considerable post-collisional
velocities in horizontal direction, in particular, if more than two
balls are involved. 
Simple but effective techniques were
introduced\cite{Harter:1971,Mellen:1968,Mellen:1995}, which allow for
experiments with chains of up to about 7 aligned balls.

In this paper we will show that although the problem looks simple, there may emerge rather complex
behavior including multiple collisions. The simple solution mentioned
above is, thus, only a special case of a more general solution.
 
\section{Independent Collision Model}
\label{sec:ICM}

\subsection{Collision Scenario}
\label{sec:collision-scenario}

To introduce our notation and for later reference let us first
reproduce the solution for the independent collisions model (ICM), that is, the
final velocity of the upper ball is the
result of a) the inelastic collision of the lower ball with the floor
and b) the inelastic collision of the lower ball with the upper
ball. Note that here and in the following we restrict ourselves to the
situation where $m_1 > m_2$. For the opposite case, $m_1<m_2$, it may
be shown that the sketched sequence of collisions fails, even under
the assumption of perfectly elastic collisions\cite{Harter:1971,Patricio:2004}.

Inelasticity of the balls is described by the coefficient of restitution
which relates the precollisional relative velocity
$v_{ij}=v_i-v_j$ of colliding particles, $i$ and $j$ to
the post-collisional one, $v_{ij}^\prime=v_i^\prime-v_j^\prime$, 
\begin{equation}
  \label{eq:cor}
  \varepsilon=\left.-v_{ij}^\prime \right/v_{ij}\,.
\end{equation}
The lower sphere ($m_1$) reaches the floor at time
$t_{10}=\sqrt{\frac{2}{g}(z_{1}^{(0)}-R_1)}$ at the velocity
$v_{10}=-\sqrt{2g(z_{1}^{(0)}-R_1)}$ from where it is reflected with $v_{10}^\prime=-\varepsilon
v_{10}$. The first index always denotes the considered particle and the second
its collision-partner. Index 0 stands for the floor, 1 for the lower and 2 for
the upper sphere. Upper index $^{(0)}$ stands for initial values.
Post-collisional values are marked by primes. The particles collide then at
$t_{12}=t_{10}-\frac{\Delta h}{v_{10}(1+\varepsilon)}\equiv t_{10}+\Delta t$
whereby the upper sphere is located at position
$z_{21}=z_{21}^\prime=-\frac{1}{2}gt_{12}^2+z_{2}^{(0)}$ with the initial height $z_{2}^{(0)}=z_{1}^{(0)}+R_1+R_2+\Delta h$. At $t_{12}$ the
lower particle has the velocity $v_{12}=-g\Delta t+v_{10}^{\prime}$  and the upper $v_{21}=-gt_{12}$. Employing the
collision rule, Eq. \eqref{eq:cor}, and conservation of momentum, we obtain the final
velocities
\begin{eqnarray}
v_{2}^\prime &=& \frac{m_2v_{21}+m_1[v_{12}+\varepsilon(v_{12}-v_{21})]}{m_1+m_2}
\label{eq:v2FinEmd}\\
v_{1}^\prime&=&\frac{m_1v_{12}+m_2[v_{21}+\varepsilon(v_{21}-v_{12})]}{m_1+m_2}
\label{eq:v1FinEmd}
\end{eqnarray}
and the relative velocity 
\begin{equation}
  \label{eq:vrelprime}
v_r^\prime=v_2^\prime-v_1^\prime\,.  
\end{equation}
In the case of elastic ($\varepsilon=1$) and instantaneous ($\Delta h\to 0$)
collisions we find
\begin{eqnarray}
\label{eq:v2p}
v_{2}^\prime&=&-v_{10}\frac{3-\mu}{1+\mu}\qquad\text{with}\quad \quad\mu=\frac{m_2}{m_1}\\
  \label{eq:1}
  v_{1}^\prime&=&-v_{10}\frac{1-3\mu}{1+\mu}\,.
\end{eqnarray}
For $\mu=\nicefrac{1}{3}$ the lower sphere loses all its kinetic energy
($v_1^\prime=0$) and the upper sphere rebounds with twice its initial velocity.
In the limit $\mu\rightarrow0$ we recover the well known textbook result
$v_2^\prime=-3v_{10}$, that is, the upper ball rises to about nine times of its
initial dropheight ($z_2^{\text{max}}=9z_{2}^{(0)}-8(2R_1+R_2)$).

The system of bouncing balls can be exhaustively described also for 
more than two spheres \cite{Kerwin:1972}, provided
the collisions are considered as isolated events, that is, only
two-particle interactions are taken into account.

\subsection{Coefficient of Restitution Resulting from the Solution of Newton's Equation}
\label{sec:coeff-rest-result}

The contact of viscoelastic spheres is described by the (modified) Hertz
contact law \cite{BrilliantovEtAl:1996}. In this article we will use a
simplified force law since it allows for a exhaustive analytical solution of the
problem. To justify this approximation, we will show later by means of numerical
simulations that the more correct Hertz contact force leads to qualitatively
identical results, see Appendix \ref{sec:AppVisco}.

We describe the contact of dissipatively interacting particles, $i$ and $j$, by
\begin{equation}
  \label{eq:dashpot}
  F\left(\xi_{ij},\dot{\xi}_{ij}\right)=\min\left[0,-k\xi_{ij}-\gamma\dot{\xi}_{ij}\right]
\end{equation}
as a function of the mutual compression
\begin{equation}
  \label{eq:xi}
  \xi_{ij}(t)=\max\left[0,R_i+R_j-\abs{\vec{r_i}(t)-\vec{r_j}(t)}\right]
\end{equation}
and the compression rate
$\dot{\xi}_{ij}=\text{d}\xi_{ij}(t)/\text{d}t$, where $R_i$ is the
radius of particle $i$ and $\vec{r}_i(t)$ is its position at time $t$. The
expression in square
brackets in Eq. \eqref{eq:dashpot} may become positive during the expansion phase,
that is, the
(positive) dissipative force may overcompensate the (negative) elastic
force which would lead to a resulting erroneous attractive
force, see e.g. \cite{SchwagerPoeschel:2008}. Therefore, the $\min[\dots]$
function is applied to take into account that the interaction force is always
repulsive (negative). 

Consider an isolated pair of colliding particles $i$ and $j$ approaching
one another at impact rate $v=\dot{\xi}(t=0)$ at $t=0$. Using the force, Eq.
\eqref{eq:dashpot}, we obtain the relative velocity after a collision by solving Newton's equation of motion, 
\begin{equation}
  \label{eq:Newton}
 m_{ij}^\text{eff}\ddot{\xi_{ij}}
=F\left(\dot{\xi_{ij}},\xi_{ij}\right)\,,
\end{equation}
with the effective mass $m_{ij}^\text{eff}=m_im_j/\left(m_i+m_j\right)$ and
initial conditions $\xi_{ij}(0)=0$ and
$\dot{\xi_{ij}}(0)=v$. The collision is complete at time $t_c$
when $\ddot{\xi}_{ij}(t_c)=0$ \cite{SchwagerPoeschel:2008}.  

Of course, for a pairwise collision the final velocity as obtained
from Eq. \eqref{eq:cor} must coincide with the final velocity as
obtained from integrating Newton's equation of motion. Therefore, the solution
$\dot{\xi}_{ij}(t_c)$ of Eq. \eqref{eq:Newton} allows to relate
the coefficient of restitution $\varepsilon$ to the parameters $k$ and $\gamma$ of the force law,
Eq. \eqref{eq:dashpot}, via
\begin{equation}
  \label{eq:cor2}
  \varepsilon = -\frac{\dot{\xi}(t_c)}{v}\,.
\end{equation}
Straightforward calculation \cite{SchwagerPoeschel:2008} yields for the
duration of the collision\\

\begin{equation}
  \label{eq:tc}
    t_c = \begin{cases}
   \displaystyle\frac{1}{\omega}\left(\pi
-\arctan\displaystyle\frac{2\beta\omega}{\omega^2-\beta^2}\right) &
\mbox{for~~~} \displaystyle\beta<\frac{\omega_0}{\sqrt{2}} \\[0.5cm]
\displaystyle-\frac{1}{\omega}\arctan\displaystyle\frac{2\beta\omega}{
\omega^2-\beta^2} & \mbox{for~~~} \displaystyle\beta>\frac{\omega_0}{\sqrt{2}}
\end{cases}
\end{equation}
with 
\begin{equation}
 \omega_0^2\equiv\frac{k}{m^{\text{eff}}}\,;~~~~~\beta\equiv
  \frac{\gamma}{2m^{\text{eff}}}\,;~~~~\omega\equiv\sqrt{\omega_0^2-\beta^2}\,.
\end{equation}
For the coefficient of restitution we obtain
\begin{equation}
  \varepsilon = \begin{cases}
    \displaystyle\exp\left[-\frac{\beta}{\omega}\left(\pi - \arctan\displaystyle\frac{2\beta\omega}{\omega^2-\beta^2}\right)\right] & \!\!\!\!\!\!\mbox{for~}  \displaystyle\beta<\frac{\omega_0}{\sqrt{2}} \\[0.3cm]
    \displaystyle\exp\left[\frac{\beta}{\omega}\arctan\displaystyle\frac{2\beta\omega}{\omega^2-\beta^2}\right]
    & \!\!\!\!\!\!\mbox{for~} \displaystyle \beta\in\left[\frac{\omega_0}{\sqrt{2}},\omega_0\right]
\\[0.3cm]
    \displaystyle\exp\left[-\frac{\beta}{\Omega}\ln\frac{\beta+\Omega}{\beta-\Omega}\right] & \!\!\!\!\!\!\mbox{for~} \beta>\omega_0
    \end{cases}
  \label{eq:dashpotCOR1}
\end{equation}
where $\Omega\equiv\sqrt{\beta^2-\omega_o^2}$.
Note that $\varepsilon$
depends on the parameters of the force
and the effective mass $m^\text{eff}$ of the colliding particles, that is,
$\varepsilon=\varepsilon(k,\gamma, m^\text{eff})$. Thus, $\varepsilon$ may not
be considered as a pure material constant.

\section{Simultaneous Contacts}
\label{sec:simult-cont}

\subsection{Equations of Motion}
\label{sec:equations-motion}

The ICM fails if we take into account the finite
duration of the collisions. In this case, it may happen that the
collision of the particles (process b) starts yet before the collision
of the lower particle with the floor (process a) has terminated. In
this case we have a three-particle interaction of the floor and both
balls which cannot be resolved using the concept of the coefficient of
restitution. In this case, the final velocity of the upper particle must be
determined by integrating Newton's equation of motion for the
three-particle system which requires the detailed knowledge of the
interaction forces. Consequently, we have to solve the set of
Newton's equations
\begin{equation}
  \label{eq:set}
  \begin{split}
    m_1\ddot{z}_1+m_1g+F_{12}-F_{01}&= 0\\
    m_2\ddot{z}_2+m_2g-F_{12}&= 0\,,
  \end{split}
\end{equation}
where $F_{ij}$ is the model-specific interaction law between
particles $i$ and $j$ and the floor is considered as particle 0 (with $m_0\to\infty$).

The failure of the simplifying ICM was discussed in the context of the
closely related problem of Newton's cradle. A simple
analysis reveals immediately that the textbook-like explanation using
isolated collisions is insufficient \cite{Kline:1960}. Instead, the details of the
interaction force must be taken into account. The explanation of
Newton's cradle is far from being simple and there is an intensive and
controversial discussion about this seemingly simple classroom experiment \cite{Chapman:1960,HerrmannSchmaelzle:1981,HerrmannSeitz:1982,PiquetteWu:1982,HerrmannSchmaelzle:1984,Reinsch:1994,HutzlerEtAl:2004,HinchSaintJean:1999}.

The necessity of considering the details of the interaction force
becomes obvious immediately when considering colliding rods instead of
spheres \cite{Auerbach:1994,Maecker:1953,FuPaul:1970}. In fact, the investigation
of longitudinal waves in colliding bodies and the corresponding duration of
the collision is a classical problem of mechanics, investigated by
some of the most eminent scientists, such as
Poisson \cite{Poisson:1833}, Boltzmann \cite{Boltzmann:1882} and other
important scientists \cite{Voigt:1915,Schneebeli:1871,Hamburger:1886}

\subsection{Comparison with the ICM}
\label{sec:comparison}

In Sec. \ref{sec:coeff-rest-result} we conclude the coefficient of restitution
from the interaction force. Using this result, we can compute the final relative
velocity $v_r^\prime$ by means of Eq. \eqref{eq:vrelprime}, employing the
assumption of independent collisions. Alternatively we can obtain $v_r^\prime$
by solving the set of equations \eqref{eq:set}
numerically. The latter approach does not require any assumption on the sequence
of the collisions. We will see that both results may deviate considerably
according to rather complex dynamics of the system.

In order to compare both results by means of Eq. \eqref{eq:dashpotCOR1} we map
the constants of the force law to the coefficient of restitution
$(k,\gamma,m^\text{eff}) \leftrightarrow \varepsilon$. 

We assume that the collisions between the lower sphere and the floor
($ij=01$) and between the spheres ($ij=12$) take place at the same coefficient
of restitution. Since the effective mass enters Eq. \eqref{eq:dashpotCOR1}, for
given material stiffness, $k=\text{const.}$, Eq. \eqref{eq:dashpotCOR1} then
provides a relation between $\varepsilon$ and $\gamma_{ij}$, thus, we can
determine $\gamma_{ij}$ by specifying $\varepsilon$ as a control parameter of
the problem.

The latter assumption implies the somewhat unphysical fact that the lower side
of the large sphere (where it contacts the floor) is characterized by a larger
dissipative constant $\gamma_{01} \ne \gamma_{12}$ than its upper side (where it
contacts the smaller sphere).  We will justify this assumption in App.
\ref{sec:same-material}  where we show that the perhaps more plausible
assumption $\gamma_{01}=\gamma_{12}$, implying $\varepsilon_{01}\ne
\varepsilon_{12}$,  leads to qualitatively identical results.

\section{Basketball -- Tennis Ball Problem}
\label{sec:tbp_lin_dash}
\subsection{Collision Sequence}
\label{sec:collision-sequence}
\begin{figure}[htbp]
  \centering
 \includegraphics[width=0.99\columnwidth,clip]{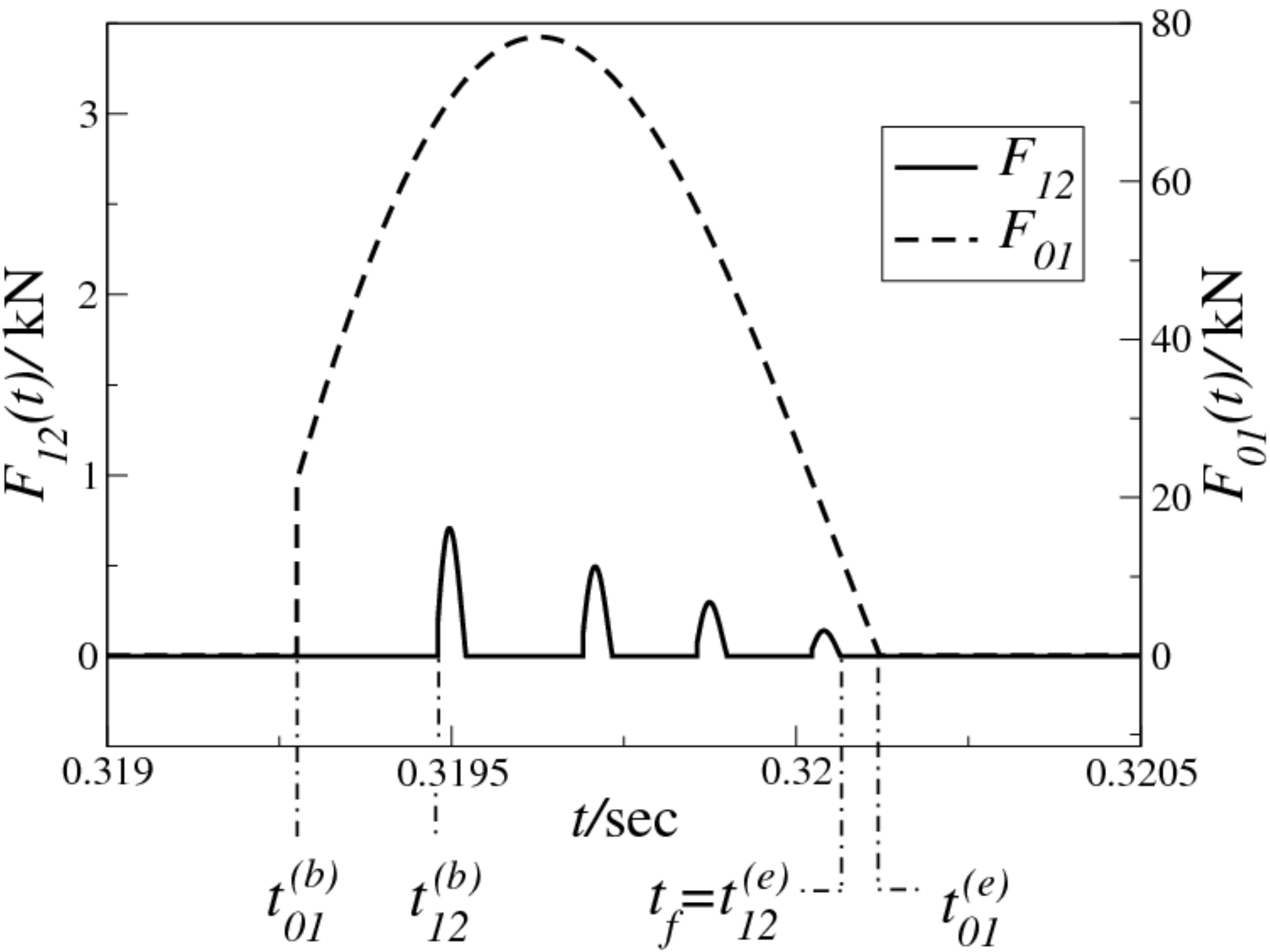}
 \caption{Forces $F_{01}$ and $F_{12}$ obtained from solving Eqs. \eqref{eq:set}.
During the contact between the lower particle and the floor there occur
multiple contacts between the spheres (full line). For discussion see the text.
Parameters: $R_1=10$\,cm, $R_2=1$\,cm, $\Delta h=0.1$\,mm, $z_{1}^{(0)}=0.6$\,m,
$k=5.0\cdot10^7$\,N/m, $\varepsilon=0.7$ (corresponding to hard rubber).
}
  \label{fig:dash_forces}
\end{figure}

Let us assume two vertically aligned balls (the basketball -- tennis ball problem) 
as sketched in Fig. \ref{fig:sketch}. We integrate Newton's equation of motion,
Eq. \eqref{eq:set}, for this system numerically and obtain the forces between
the bottom and the lower sphere and between both spheres, see
Fig. \ref{fig:dash_forces}.
For time  $t_{01}^\text{(b)} \le t \le t_{01}^\text{(e)}$ the lower particle is
in contact with the floor as indicated by the force $F_{01}\ne 0$. During this
interval the balls are in contact repeatedly, starting at time (first contact) 
$t_{12}^\text{(b)}$ and ending (last contact) at time $t_{12}^\text{(e)}$ as
indicated by $F_{12}\ne0$. An interesting detail is the discontuity of $F_{01}$ at $t=t_{01}^{(b)}$ which is a consequence of the force law, Eq. \eqref{eq:dashpot}: At the instant of the contact where $\xi_{01}\to 0$, the elastically restoring term, $k\xi_{01}$, vanishes whereas the (repulsive) dissipative term, $\gamma\dot{\xi}_{01}$, has a finite value as soon as the particles get into contact. 


The existence of multiple collisions shown in Fig. \ref{fig:dash_forces} shows
that the ICM described in Sec. \ref{sec:ICM} fails for the chosen set of
parameters which provokes mainly two questions:
\begin{enumerate}
\item How many contacts between the spheres occur and how does their number
depend on the system parameters ($\Delta h$, $R_1$, $R_2$, $\gamma_{ij}$ or
$\varepsilon$ respectively)?
\item If multiple collisions take place, when does the collision sequence terminate?  
\end{enumerate}
Depending on the system parameters we may obtain
$t_{12}^\text{(e)}\leq t_{01}^\text{(e)} $ or
$t_{12}^\text{(e)}>t_{01}^\text{(e)} $, therefore, the second question must be
answered by a definition: The collision-sequence terminates at time $t=t_f$ (see
Fig. \ref{fig:dash_forces}) when the last contact between the spheres ceases, before the large sphere collides with the floor for the
second time.

\begin{figure}[htbp]
  \centering
  \includegraphics[width=0.99\columnwidth]{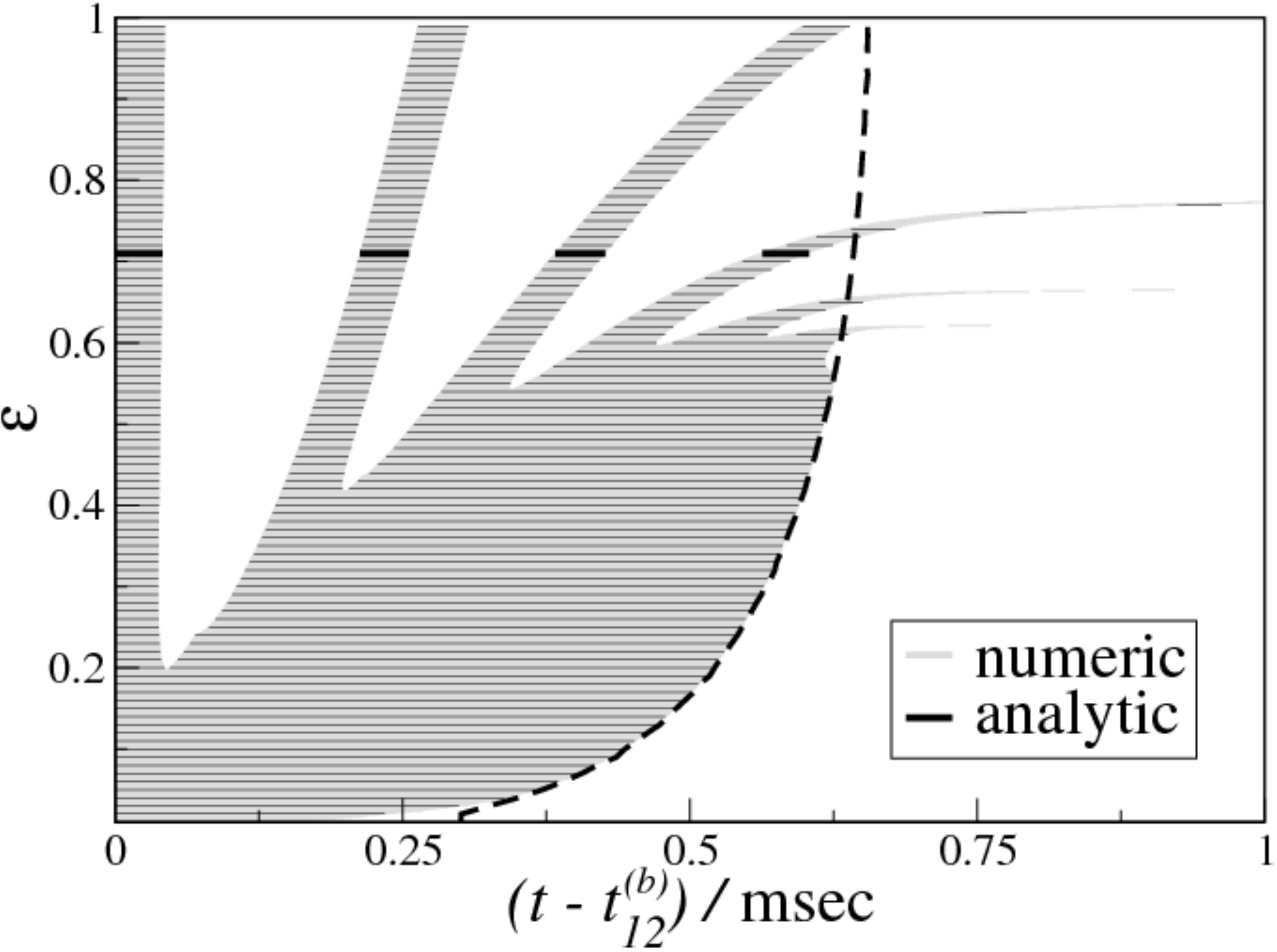}
\caption{Sequence of collisions for varying coefficient of restitution. The
dashed line shows the end of the contact between the lower sphere and the floor,
$t_{01}^\text{(e)}$. 
The fat line corresponds to the force drawn in Fig. \ref{fig:dash_forces}
($R_1=10$ cm, $R_2=1$ cm, $\Delta h=0.1$ mm, $z_{1}^{(0)}=0.6$ m,
$k=5.0\cdot10^7$ N/m).}
  \label{fig:dash_contacts}
\end{figure}

To answer the first question,
we refer to Fig. \ref{fig:dash_contacts} which illustrates the sequence of
collisions in dependence of the coefficient of restitution $\varepsilon$ for
fixed $\Delta h$ and $R_1/R_2$.
The value of $\varepsilon$ was adjusted by varying $\gamma$, according to
Eq. \eqref{eq:dashpotCOR1} while keeping $k=5.0\cdot10^7\text{N}/\text{m}$ invariant.
Figure \ref{fig:dash_contacts} should be read horizontally (for fixed value of
$\varepsilon$): each black or grey line marks time intervals when the particles are in
contact. 
For elastic balls, $\varepsilon=1$, and the chosen parameters there occur 3
collisions. 
For sufficiently large $\Delta h$, this number may be unity, that is, the
independent-collisions condition is fulfilled (see below). Keeping $\Delta h$, $k$
and the radii $R_1$ and $R_2$ constant and decreasing $\varepsilon$, the number
of contacts increases. This is due to the fact that the relative velocity of the
balls decreases because of inelastic collisions and, thus, the intervals of free
flight become shorter while the duration of the contacts depends only weakly on
the value of the inelasticity. 
For yet smaller $\varepsilon$ the relative velocity after the $k^\text{th}$
contact may be small enough such that the lower ball catches up with the upper
because of its upwards acceleration due to its contact with the floor. This
effect makes some free-flight intervals vanish for decreasing $\varepsilon$ and,
thus, reduces the number of contacts. Summarizing, for each set of parameters
$\{\Delta h$, $R_1$, $R_2$, $k\}$ the number of collisions as a function of
$\varepsilon$ is a function with a single 
maximum. 

For the force law Eq. \eqref{eq:dashpot} the basketball -- tennis ball problem may
be solved analytically by a piecewise procedure, see App. \ref{sec:ana}. To
check against numerical errors, the
horizontal gray lines (in between the black lines) show the same information as
the result of an analytical theory which agrees perfectly with the numerical
data.

There is an interesting case when the final velocity of the lower ball
after losing contact with the floor is only slightly larger than the velocity of
the upper ball after the previous collision. Since both balls move only under
the action of gravity, the balls may collide an ultimate time even after the contact between the
lower ball and the floor has already finished. These events may be seen in Fig.
\ref{fig:dash_contacts} as narrow spikes at
$\varepsilon\approx0.78$,
$\varepsilon\approx0.67$, etc. 



\begin{figure}[htbp]
  \centering
\includegraphics[width=0.99\columnwidth]{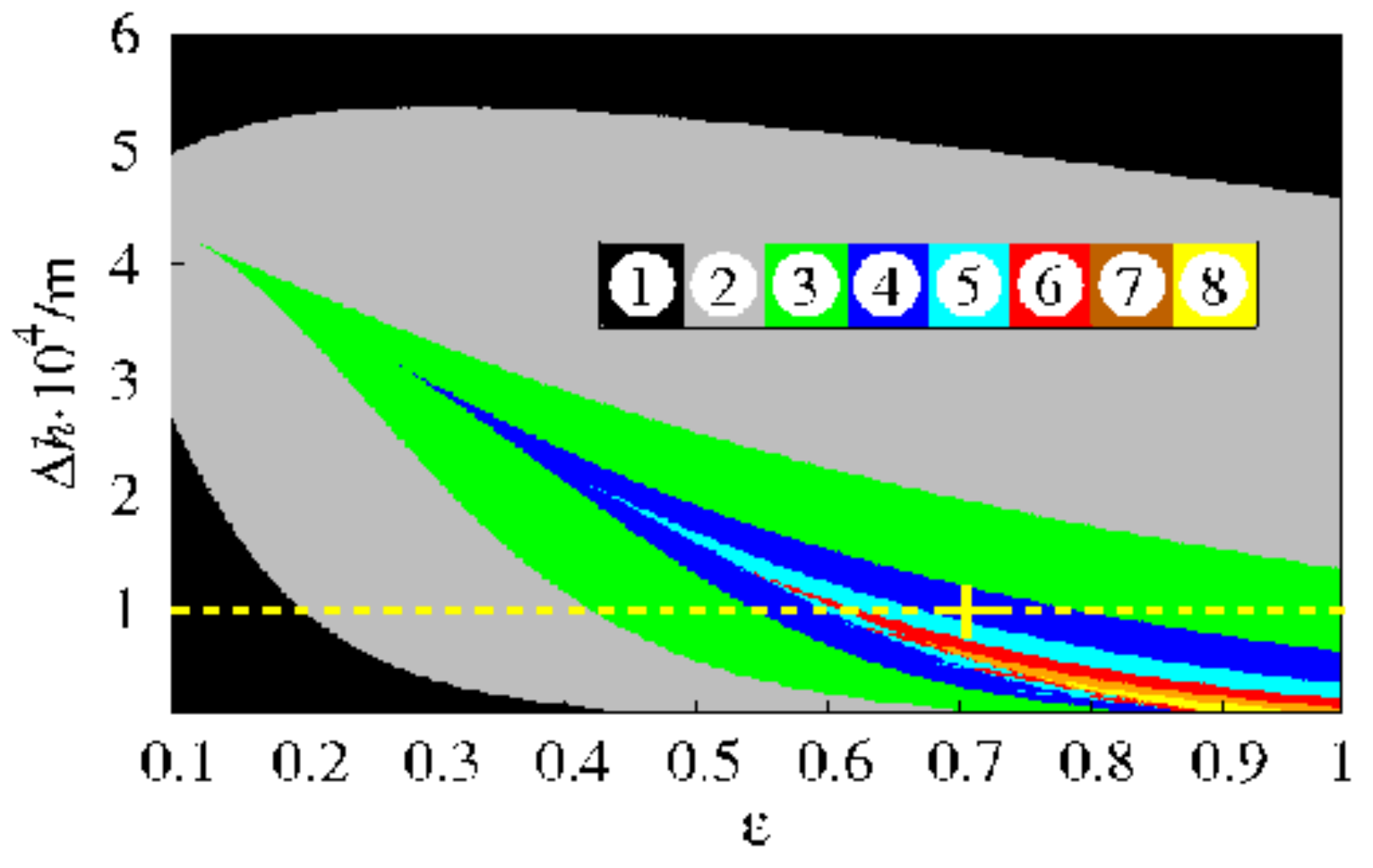}
\includegraphics[width=0.99\columnwidth]{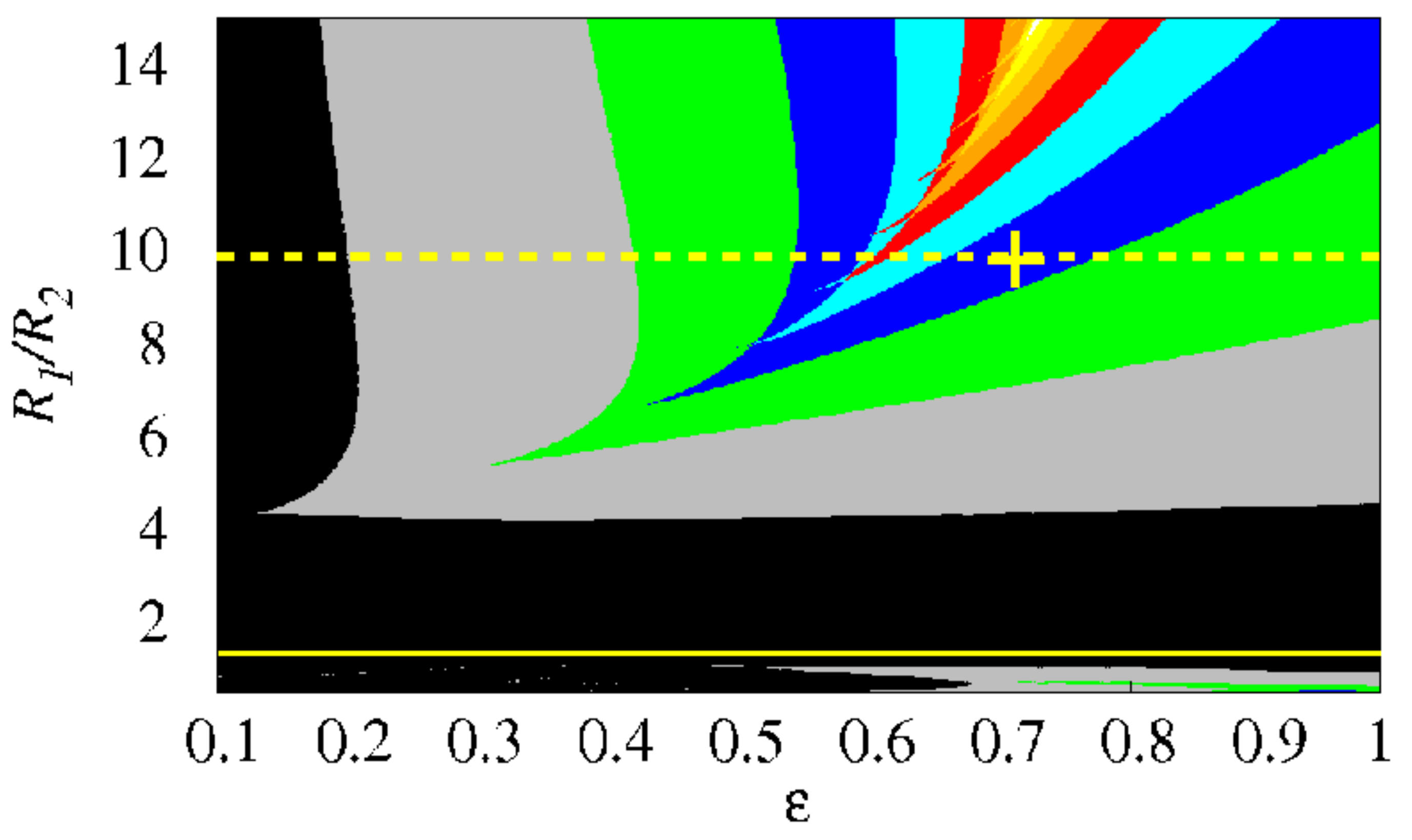}
\caption{Number of contacts between
the spheres as a function of $\varepsilon$ and $\Delta h$ (top) and
$\varepsilon$ and $R_1/R_2$ (bottom).
The dashed lines show the value of $\Delta h$ and
$R_1/R_2$ used
in Fig. \ref{fig:dash_contacts}; the $+$ symbol shows the parameters used
in Fig. \ref{fig:dash_forces}. The solid line in the lower panel indicates
$R_1/R_2=1$.  (Parameters: $k=5.0\cdot10^7\text{N/m}$,
$R_1=10\text{cm}$ and $R_2=1\text{cm}$ (top) and $k=5.0\cdot10^7\text{N/m}$,
$R_2=1\text{cm}$ and $\Delta h=0.1\text{mm}$ (bottom))}
  \label{fig:NoContacts}
\end{figure}

The number of contacts of the spheres as a function of $\varepsilon$ and the
initial distance $\Delta h$ is shown in more detail in Figure
\ref{fig:NoContacts} (top). As explained above for each value of $\varepsilon$
there is an interval for $\Delta h$ which maximizes the number of
contacts.

For the interaction force, Eq. \eqref{eq:dashpot} the ratio $t_{c,01}/t_{c,12}$ of the contact duration of the collision lower sphere/ground $t_{c,01}$ and the contact duration of the collision lower/upper sphere $t_{c,12}$ increases with $m_1/m_2$ (or $R_1/R_2$, respectively). Consequently, the number of
contacts increases with $R_1/R_2$, shown in Fig. \ref{fig:NoContacts} (bottom). On the other hand, increasing $R_1/R_2$ also increases the initial relative velocity between the two spheres and with that the intervalls of free flight, what in turn reduces the possible number of contacts. Whereas the effect explained first is dominating, the interplay of both effects explaines the rather complex behaviour shown in the bottom panel of Fig. \ref{fig:NoContacts}.




From Figs. \ref{fig:dash_contacts}  and \ref{fig:NoContacts} we see that for a
vast range of parameters the true collision scenario as obtained from the
integration of Newton's equations of motion deviates drastically from the
independent-collision scenario outlined in Sec. \ref{sec:ICM}. 

\subsection{Effective Coefficient of Restitution}
\label{sec:effCor}
By solving Newton's equation, we can compute the final relative
velocity $v_r^\prime=v_2(t_f)-v_1(t_f)$ which corresponds to Eq.
\eqref{eq:vrelprime} obtained from the ICM. To compare both results, we compute
$v_r^\prime$ by integrating Eq. \eqref{eq:set} using the interaction force Eq. \eqref{eq:dashpot}  for a certain set of
parameters  $\{\Delta h,\:m_1,\:m_2,\:\:k\}$ and a specified
$\varepsilon=\varepsilon_\text{spec}$ (which in turn determines $\gamma$ via Eq.
\eqref{eq:dashpotCOR1}). Then, by inverting Eq. \eqref{eq:vrelprime} we
determine the coefficient of restitution $\varepsilon=\varepsilon_\text{eff}$
which would yield the same final relative velocity for the ICM. If
$\varepsilon_\text{eff} /\varepsilon_\text{spec}\approx 1$, both models yield
the same result, that is, the ICM is an acceptable approximation. Otherwise, the
ICM fails.

\begin{figure}[htbp]
  \centering
   \includegraphics[width=0.99\columnwidth]{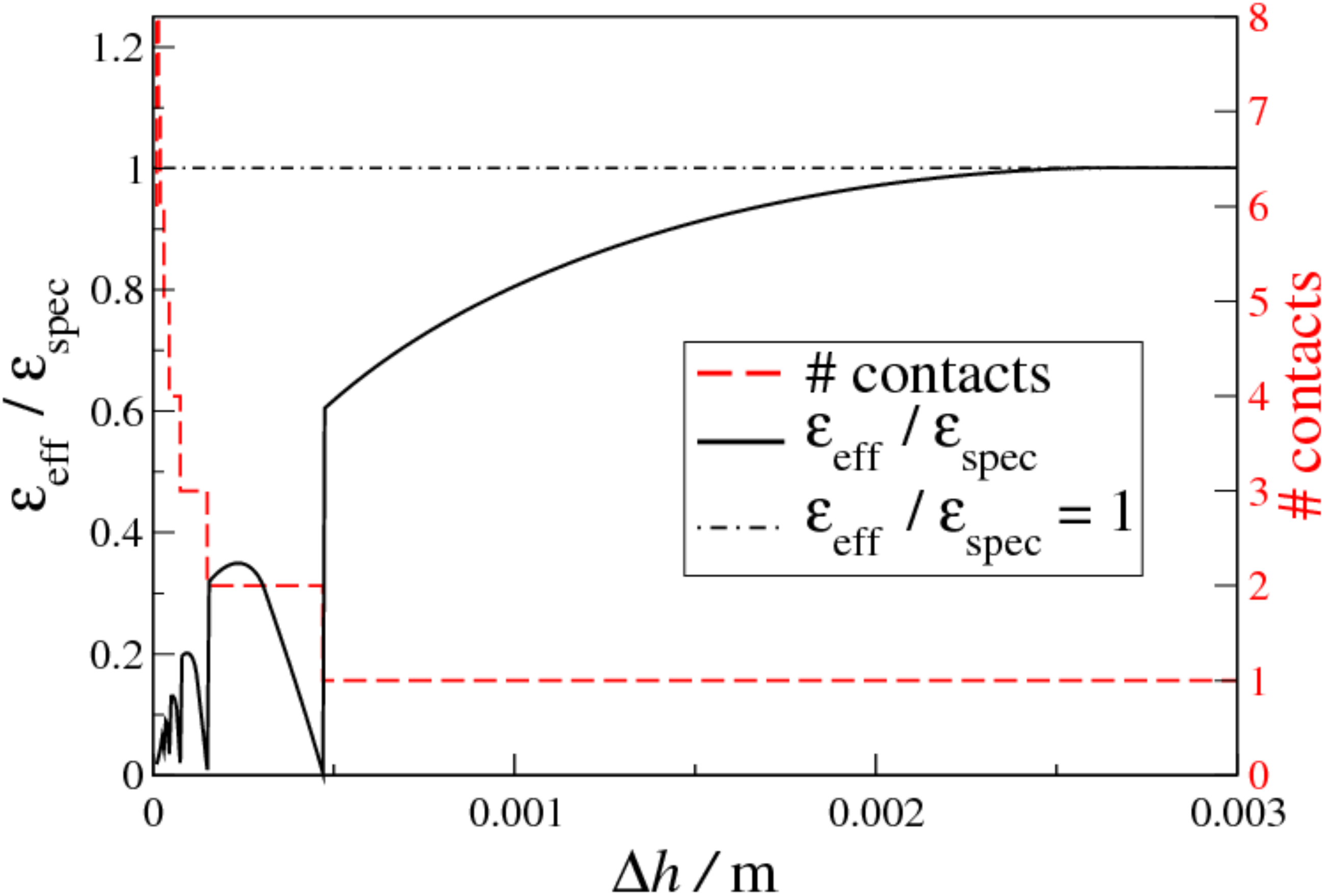}
%
\includegraphics[width=0.99\columnwidth]{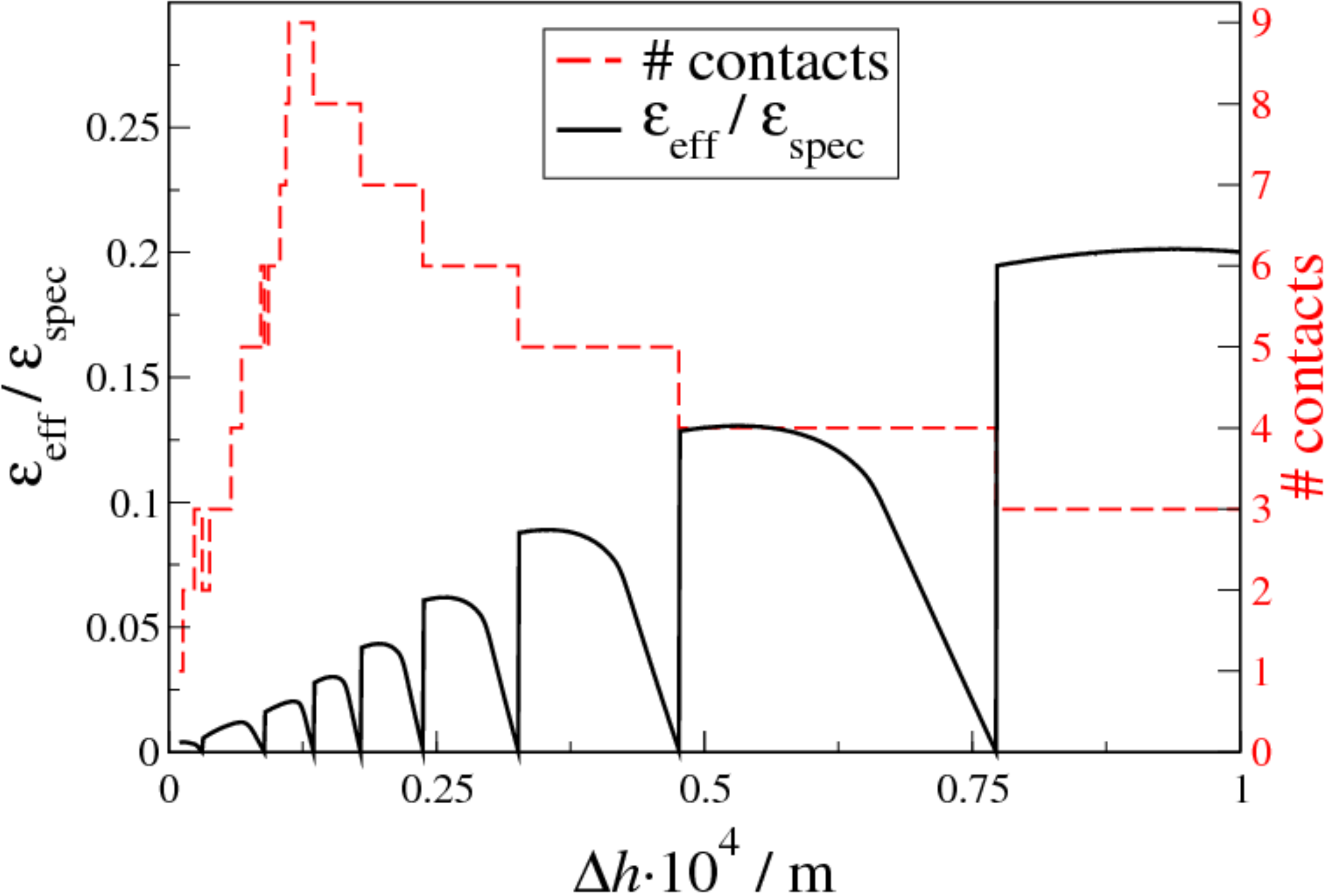}
 \caption{
$\varepsilon_\text{eff}/\varepsilon_\text{spec}$ as a function of $\Delta h$ and
the corresponding number of contacts (right axis, top). The bottom figure shows a
magnification of the small $\Delta h$ range. Parameters: 
$R_1=10\text{cm}$, 
$R_2=1\text{cm}$,  
$k=5.0\cdot10^7\text{N/m}$, 
$\varepsilon_\text{spec}=0.9$. 
}
  \label{fig:EpsRat_dh}
\end{figure}

Consider the dependence of $\varepsilon_\text{eff}/\varepsilon_\text{spec}$
on the initial distance $\Delta h$. For large $\Delta h$ the lower sphere leaves
the floor before it contacts the upper one, that is, the ICM holds true. Figure
\ref{fig:EpsRat_dh} (top) shows that
$\varepsilon_\text{eff}/\varepsilon_\text{spec} \to 1$ with increasing $\Delta
h$. Moreover, as expected for $\varepsilon_\text{eff}/\varepsilon_\text{spec}\to
1$ there is only one contact which is a necessary (but not sufficient)
precondition for independent collisions. 
Figure \ref{fig:EpsRat_dh} (bottom) is a magnification of the range of small
$\Delta h$. As discussed before, the number of contacts as a function of $\Delta h$ has
a maximum. 
The oscillations in the number of contacts as a function of $\Delta h$ for very small $\Delta h$ correspond to the spikes shown in Fig. \ref{fig:dash_contacts} where the lower sphere catches up with the upper after the lower sphere has already left the ground.


\begin{figure}[h!]
  \centering
  \includegraphics[width=0.99\columnwidth]{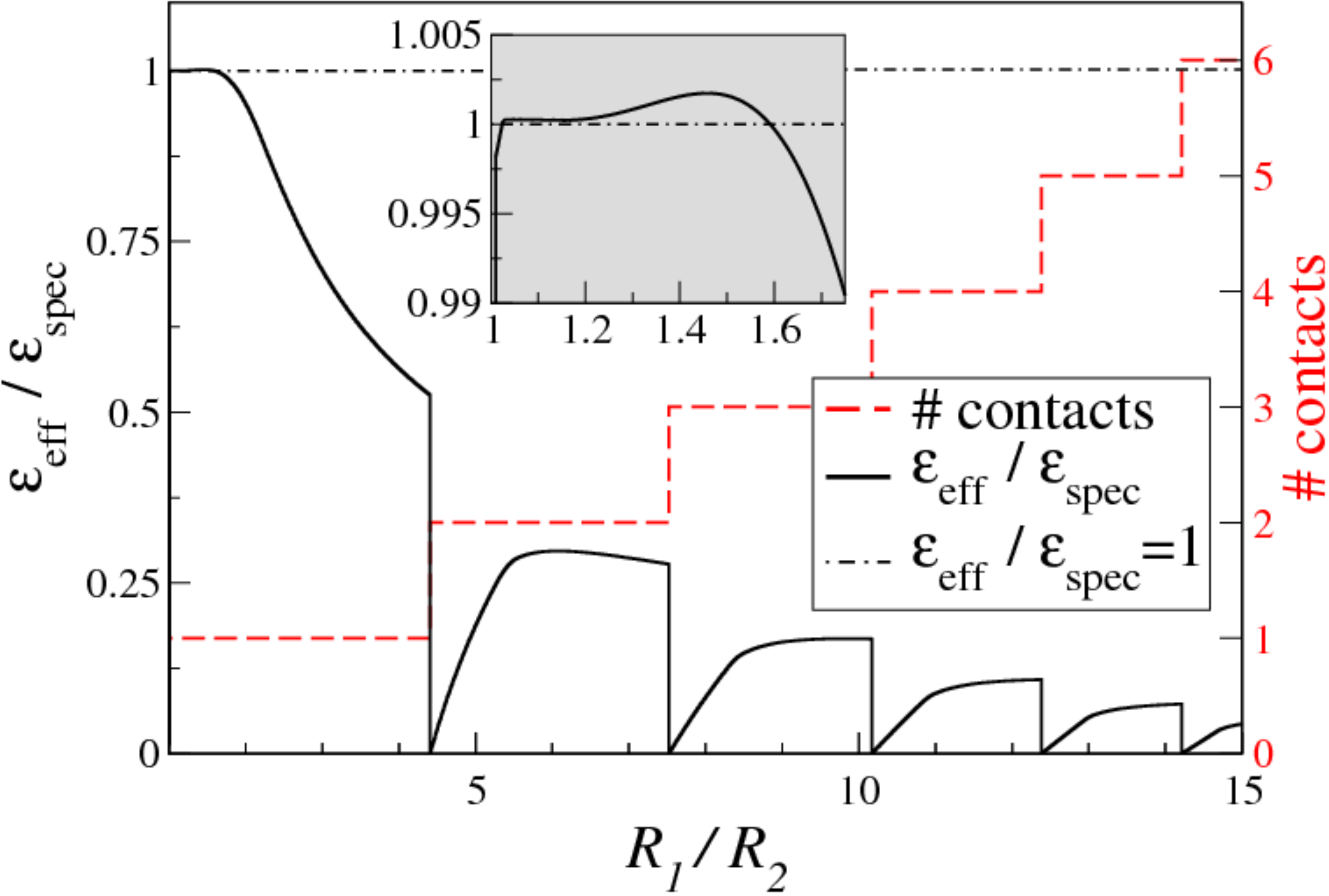}
\caption{$\varepsilon_\text{eff}/\varepsilon_\text{spec}$ as a function of
$R_1/R_2$ (with $R_2=1cm$) and the corresponding number of contacts
(right axis). Further parameters: $k=5.0\cdot10^7\text{N/m}$,
$\varepsilon_\text{spec}=0.8$, $\Delta h=0.1\text{mm}$.}
  \label{fig:EpsRat_rr}
\end{figure}
Similarly, Fig. \ref{fig:EpsRat_rr} shows $\varepsilon_\text{eff}/\varepsilon_\text{spec}$ as a function of $R_1/R_2$. Since
increasing $R_1/R_2$ increases the number of contacts, $\varepsilon_\text{eff}/\varepsilon_\text{spec}$ decreases and thus, as expected, the ICM invalidates with increasing $R_1/R_2$.



While for most values of the parameter space $\varepsilon_\text{eff}/\varepsilon_\text{spec}<1$, there is a small interval of $R_1/R_2$ where 
$\varepsilon_\text{eff}/\varepsilon_\text{spec}>1$ (inset in Fig, \ref{fig:EpsRat_rr}) which is a deviation from the ICM too. Here, because of the similar durations of the contact upper sphere/lower sphere and lower sphere/floor, the lower sphere still being in contact with floor pushes the upper one upward.  As far as we see, this is the only (tiny) effect which allows for
$\varepsilon_\text{eff} > \varepsilon_\text{spec}$.


\section{Conclusion}

We considered the motion of two vertically aligned spheres which are released to collide with the floor under the action of gravity. For the analysis of the dynamics of this textbook problem (basketball -- tennis ball problem) we used two complementary methods. First we described the system exploiting the {\em independent collision model} (ICM) which assumes instantaneous collisions between the spheres and between the lower sphere and the floor. The collisions are described by a single number, the coefficient of restitution, $\varepsilon$, and the duration of the collisions is neglected. Second, we described the dynamics by analytically and numerically solving Newton's equation of motion. Here the collisions are characterized by an interaction force law, $F(\xi,\dot{\xi})$. For the case of the linear dashpot model used here, the force is a function of the elastic and dissipative parameters $k$ and $\gamma$. Since there is a direct relation between $\varepsilon$ and $\{k,\gamma\}$, we can compare the results of the ICM and the solution of Newton's equations.  

We specify two characteristics of the process, a) the final relative velocity, $v^\prime$, between the spheres and b) the number of collisions between the spheres during the process. If the approaches were equivalent, we should obtain equivalent results for a) and b).

Obviously, in case of the ICM there is only one collision between the spheres and the final result for $v^\prime$ is the solution of a textbook problem, Eq. \ref{eq:vrelprime}. In the article we show that Newton's equations yield a different scenario, including multiple collisions and the ICM is only valid in a certain limit, that is, the ICM fails for a wide range of parameters.

To quantify the deviations, we solve Newton's equations with parameters $k$ and $\gamma$ that correspond to a certain specified coefficient of restitution $\varepsilon=\varepsilon_\text{spec}$. Then we compare this value with the {\em effective} coefficient of restitution, $\varepsilon_\text{eff}$, obtained from the final relative velocity as obtained from Newton's equation. The value $\varepsilon_\text{eff}/\varepsilon_{spec}=1$ would indicate that both models agree. Our results reveal, however, a dramatic deviation from this ideal behavior. In Figs. \ref{fig:EpsRat_dh} and \ref{fig:EpsRat_rr} we see that in contradiction to the ICM, the ratio $\varepsilon_\text{eff}/\varepsilon_{spec}$ may adopt any value from almost zero up to slightly larger than one, that is, the ICM fails dramatically.  

While our subject, the basketball -- tennis Ball problem, is only a cute but relatively unimportant toy problem, our results may have serious consequences for numerical simulation techniques of granular many-particle problems. There exist two established simulation techniques for the simulation of granular systems, Molecular Dynamics (MD) and Event-driven Molecular Dynamics (EMD). While MD solves Newton's equations of motion for all $N$ particles constituting the granular system, thus, solves a system of $3N$ (without rotation) coupled, strongly non-linear differential equations, EMD describes the dynamics of the $N$-particle system as a sequence of pairwise collisions. The latter approach allows for a great speedup of the numerical simulation since instead of solving computer-time intensive solutions of differential equations, we only have to compute postcollisional velocities from the precollisional ones, as a function of the coefficient of restitution for each pair of colliding particles, $\{\vec{v}_i, \vec{v}_j,\varepsilon\} \to \{\vec{v}_i^{\,\prime}, \vec{v}_j^{\,\prime}\}$ via a simple propagation function. In between the collisions the particles follow simple ballistic trajectories.



It is obvious, that EMD allows for very efficient simulations as compared with MD, in particular for large $N\sim 10^6\dots 10^8$, however, this speedup comes for the price of the assumption of independent collisions, that is, EMD assumes instantaneous collisions neglecting the duration of the collisions. While this assumption may be justified in a granular gas where the mean free flight time is large as compared to the typical duration of collisions, it fails for dense systems. Our simple one-dimensional, 3-particle system shows that the failure may be dramatic.

For the analytical calculations presented in this article we made two major assumptions whose justification might not be obvious beforehand: First we assumed a linear-dashpot force, Eq. \ref{eq:dashpot}, for the interaction of viscoelastic spheres. This force allows for a simple mapping of the constants $k$ and $\varepsilon$ to the coefficient of restitution which is, moreover, a constant in this case. Of course, the interaction of spheres is described by a (modified) Hertz law which leads to a impact velocity dependent $\varepsilon$. We could perform the entire calculation presented here also for the Hertz law, however, at a {\em much} larger mathematical effort (see \cite{SchwagerPoeschel:2008a} for a similar calculation). We prefer here the simplified force and demonstrate in Appendix \ref{sec:AppVisco} that the Hertz law leads to qualitatively identical results. 

The second simplification concerns the assumption of a universal coefficient of restitution for the description of the collisions between the particles and between the lower particle and the floor. Since the effective mass enters the mapping between the force constants and $\varepsilon$, the assumption of a universal $\varepsilon$ implies that the lower sphere is characterized by a certain set of parameters $\{k,\gamma\}$ when colliding with the floor, but by a different set of parameters when colliding with the upper sphere. The alternative assumption of invariant material parameters is, perhaps, more plausible but leads then to different values of the coefficient of restitution for particle-particle and particle-floor collisions. While these alternative assumptions lead, of course, to different results, in Appendix \ref{sec:same-material} we demonstrate that the qualitative properties of the dynamics are the same for both assumptions.

\appendix
\section{Analytical description}
\label{sec:ana}
For an approximative analytical description, we assume that the motion of the
large ball is not affected by the small ball. This adiabatic approximation
becomes exact for $R_1\gg R_2$ and
Eqs. \eqref{eq:set} decouple and may be solved piecewise. We obtain four different
types of collective motion:
\medskip

\noindent {\em Type A):  The balls are isolated from one another and from the floor.}
Here the particles move along ballistic trajectories
\begin{equation}
  \begin{split}
z_{1,\text{free}}(t)=-gt^2/2+v_{1}^{(k)}t+z_{1}^{(k)}\\
z_{2,\text{free}}(t)=-gt^2/2+v_{2}^{(k)}t+z_{2}^{(k)}\,.
 \end{split}
\end{equation}
In our notation $v_{i}^{(k)}$ and $z_{i}^{(k)}$ stand for the positions and velocities of the
particles at the instant when the system enters a type of motion for the $k^\text{th}$ time, that is, they are initial conditions of the piecewise analytical solution. 
\medskip

\noindent{\em Type B): The lower ball is in contact with the floor while the upper one moves freely.}
Here the upper sphere moves along a ballistic trajectory, $z_{2,\text{free}}(t)$
while the lower one moves according to a damped harmonic oscillator,
\begin{equation}
m_1\ddot{z}_1=-m_1g+k_1(R_1-z_1)-\gamma_1\dot{z}_1
\end{equation}
with the solution
\begin{equation}
\label{eq:typeb}
z_{1,\text{ground}}(t)=A\cos[\omega t+p]\mathrm e^{-\lambda t}+z_{1,\text{inh}}.
\end{equation}
where
\begin{equation*}
  \begin{split}
&p=\arctan\left[\frac{\lambda(z_{1}^{(k)}-z_{1,{\rm inh}})+v_{1}^{(k)}}{(z_{1,\text{inh}}-z_{1}^{(k)})\omega}\right]\\
&A=\frac{z_{1}^{(k)}-z_{1,\text{inh}}}{\cos(p)}\,;~~~z_{1,\text{inh}}=\frac{k_1R_1-m_1g}{k_1}\\
&  \lambda=\frac{\gamma_1}{2m_1}\,;~~~
\omega=\sqrt{\omega_0^2-\lambda^2}\,;~~~\omega_0=\sqrt{k_1/m_1}\,.
     \end{split}
\end{equation*}
\medskip

\noindent{\em Type C):
The balls are in contact and the lower ball contacts the floor.}
Here the lower sphere moves due to Eq. \eqref{eq:typeb}, disregarding the force
resulting from the contact with the upper sphere (adiabatic approximation). The
latter moves like a damped harmonic oscillator in the presence of gravity,
additionally driven by the motion $z_{1,\text{ground}}(t)$ of the lower sphere:
\begin{multline}
\label{eq:DGLz2ground}
m_2\ddot{z}_2=-m_2g+k_2[R_1+R_2-(z_2-z_{1,\text{ground}})]\\
-\gamma_2(\dot{z}_2-\dot{z}_{1,\text{ground}}).
\end{multline}
The solution $z_{2,\text{ground}}(t)$ of Eq. \eqref{eq:DGLz2ground} is
straightforward and similar to Eq. \eqref{eq:typeb} but since it is rather
lengthy it is not given here. 
\medskip

\noindent {\em Type D): The balls are in contact with each other but not with the floor.}
Here the lower sphere follows a ballistic trajectory disregarding the force
resulting form the contact with the upper one (adiabatic approximation) and the
upper sphere moves as described in {\em type C)} but now driven by the motion
$z_{1,\text{free}}(t)$ of the lower sphere:
\begin{multline}
\label{eq:DGLz2air}
m_2\ddot{z}_2=-m_2g+k_2[R_1+R_2-(z_2-z_{1, \text{free}})]\\
-\gamma_2(\dot{z}_2-\dot{z}_{1, \text{free}})\,.
\end{multline}
Again we do not provide the lengthy but straightforward solution $z_{2,
\text{air}}(t)$ of Eq. \eqref{eq:DGLz2air} here.

We keep in mind that balls $i$ and $j$ (with $i=0$ representing the floor) are
in contact if the mutual compression $\xi_{ij}$ is positive and the interaction force
$F_{ij}$ is repulsive. Then we obtain the analytical solution of the problem,
$z_1(t)$ and $z_2(t)$, from combining the analytical solutions of the cases {\em A-D}
by means of the following scheme:
\begin{enumerate}
\item {\em Type A} motion until the lower sphere touches the floor at
$T_\text{bcl}$ where $(\xi_{01}(T_\text{bcl})>0)\wedge(F_{01}(T_\text{bcl})>0)$.
\item {\em Type B} motion until the spheres contact each other at $T_\text{bcu}$
where $(\xi_{12}(T_\text{bcu})>0)\wedge(F_{12}(T_\text{bcu})>0)$.
\item {\em Type C} motion until the contact between the spheres breaks at
$T_\text{ecu}$ where $F_{12}(T_\text{ecu})\leq0$.
\item Repeat steps 2 and 3 until the lower sphere leaves the floor at $T_{ecl}$
where $F_{01}(T_{ecl})\leq0$:\\ 
If $T_\text{bcu}>T_{ecl}$ $\to$ {\em Type B} motion until $T_{ecl}$.\\
If $T_\text{ecu}>T_{ecl}$ $\to$ {\em Type C} motion until $T_{ecl}$ and then
{\em Type D} motion until the spheres separate at $T_\text{ecu}$ where
$F_{12}(T_\text{ecu})\leq0$.
\item {\em Type A} motion until
\begin{enumerate}
\item The lower sphere contacts the floor for the second time at
$T_\text{bcl}$ where $(\xi_{01}(T_\text{bcl})>0)\wedge(F_{01}(T_\text{bcl})>0)$,
or
\item The spheres touch each other again at $T_\text{bcu}$
where $(\xi_{12}(T_\text{bcu})>0)\wedge(F_{12}(T_\text{bcu})>0)$.
\end{enumerate}
In the first case the collision sequence has terminated. In the second case:
{\em Type D} motion until the spheres separate at $T_\text{ecu}$ where
$F_{12}(T_\text{ecu})\leq0$ or until the lower sphere contacts the ground at
$T_\text{bcl}$ where $(\xi_{01}(T_\text{bcl})>0)\wedge(F_{01}(T_\text{bcl})>0)$.
\end{enumerate}

The described procedure seems to be circumstantial but it provides an
exact analytical solution of the problem in adiabatic approximation.

\section{Validity of the Simplified Force Model}
\label{sec:AppVisco}

The purpose of this Appendix is to demonstrate that the analytical and
numerical results presented in Sec. \ref{sec:tbp_lin_dash} are more than just
artifacts of the simplified interaction force Eq. \eqref{eq:dashpot}. The reason
of the deviation of the {\em effective} coefficient of restitution from the {\em
specified} coefficient shown in Figs. \ref{fig:EpsRat_dh} and
\ref{fig:EpsRat_rr} are the described multiple collisions arising from the
finite duration of the collisions. Therefore, here we show that multiple
collisions also appear for the much more realistic interaction force Eq.
\eqref{eq:visc_el_force}. To this end we reproduce Fig. \ref{fig:dash_contacts}
where the time intervals of particle contacts are shown in dependence of the
(specified) coefficient of restitution.

\subsection{Viscoelastic Spheres}
\label{sec:visco}

The force law, Eq. \eqref{eq:dashpot}, is convenient as it leads to a
coefficient of restitution, Eq. \eqref{eq:dashpotCOR1}, in an elementary way. However, this force
law is a strong simplification. Perhaps the simplest particle interaction
model which is not in conflict with basic mechanics of materials, is the contact
of viscoelastic
spheres, $i$ and $j$ \cite{BrilliantovEtAl:1996}, given by
\begin{equation}
  \label{eq:visc_el_force}
 \frac{F(\xi_{ij},\dot\xi_{ij})}{m_{{ij}}^{\text{eff}}}=\min\left[0,-k\xi_{ij}^{3/2} -
\gamma\sqrt{\xi_{ij}}\dot\xi_{ij}\right]\,,
\end{equation}
with
\begin{equation}
\rho\equiv\frac{2Y\sqrt{R_{ij}^\text{eff}}}{
3\left(1-\nu^2\right)},\text{~~}k\equiv\frac{\rho}{m_{{ij}}^{\text{eff}}},\text{~~}
\gamma\equiv\frac{3}{2}\frac{\rho A}{m_{{ij}}^{\text{eff}}},
\end{equation}
the Young modulus $Y$, the Poisson ratio $\nu$, the effective
radius $R_{ij}^\text{eff}=R_i R_j /\left(R_i+R_j\right)$ and the effective
mass $m_{ij}^\text{eff}=m_i m_j /\left(m_i+m_j\right)$. Again we use the mutual compression
and the compression rate to describe the contact dynamics (see Eq. \eqref{eq:xi}).
The dissipative constant $A$ is a function of the material viscosity;
see \cite{BrilliantovEtAl:1996} for details. For {\em elastic} spheres,
$A=0$, we recover the classical Hertz contact force \cite{Hertz:1882}.

A necessary prerequisite for deriving Hertz's law of contact and
its generalization to viscoelastic spheres, Eq. \eqref{eq:visc_el_force},  is
the assumption of small particle deformation, that is, the interaction force
causes only local displacements in the region of the contact area. Moreover, the
impact rate must be small as compared to the speed of sound to allow for a
quasistatic approximation, see \cite{BrilliantovEtAl:1996}. More complex
deformations including surface waves and oscillations, e.g. \cite{Cross:1999},
are not considered here. Such oscillations may also give rise to multiple
collisions between particles and yet more complicated particle-particle
interaction.

The relation between the coefficient of restitution and the parameters of the
force law, corresponding to Eq. \eqref{eq:dashpotCOR1}, may also be
obtained for the case of viscoelastic spheres. The calculation is
cumbersome \cite{GGcool_98,SchwagerPoeschel:2008a} (a simplified version
 is based on a dimension analysis \cite{GranRospap_99}), here we present only the
result:
\begin{equation}
  \label{eq:CORvisco}
 \varepsilon(v)=\sum_{k=0}^\infty h_k\beta^{k/2}v^{k/10}\,,
\end{equation}
with the initial conditions $\xi_{{ij}}(0)=0$, $\dot\xi_{{ij}}(0)=v$ and with $\beta\equiv\gamma k^{-3/5}$ and the pure numbers $h_0=1$, $h_1=0$, $h_2=-1.153$, $h_3=0$, $h_4=0.798$, $h_5=0.267$, \dots (see \cite{SchwagerPoeschel:2008a} for the numerical values $h_k$).
Note that in contrast to the previous case,
Eq. \eqref{eq:dashpotCOR1}, here the coefficient of restitution
depends on the impact velocity $v$.

\subsection{Basketball -- Tennis Ball Problem for  Viscoelastic Balls}
\label{sec:tbp_visco}

Just as in Sec. \ref{sec:tbp_lin_dash} we use the coefficient of
restitution $\varepsilon$ to characterize the system's dissipative properties,
due to the dissipative constant $A$ in the force law, Eq.
\eqref{eq:visc_el_force}. We proceed on the lines of Sec.
\ref{sec:collision-sequence}: We specify $\varepsilon$, the Young modulus $Y$,
the poisson ratio $\nu$, the effective radius $R_{ij}^\text{eff}$ and mass
$m_{ij}^\text{eff}$ and solve Eq. \eqref{eq:CORvisco} numerically for the
dissipative parameter $A$. Additionally, we specify the initial velocity
$v_{in}=3$ m/s since for the viscoelastic force law the coefficient of
restitution depends on the impact rate. 
As in Sec. \ref{sec:tbp_lin_dash}, the assumption of a universal value of the
coefficient of restitution to describe both particle-particle and particle-floor
contact results in the rather artificial fact that the spheres cannot consist of
the same material. This assumption is necessary to use the quite descriptive
quantity $\varepsilon$ as a control parameter. In App. \ref{sec:same-material}
we will release this assumption and show that it does not qualitative change the system's
behavior.

\begin{figure}[htbp]
  \centering
  \includegraphics[width=0.99\columnwidth]{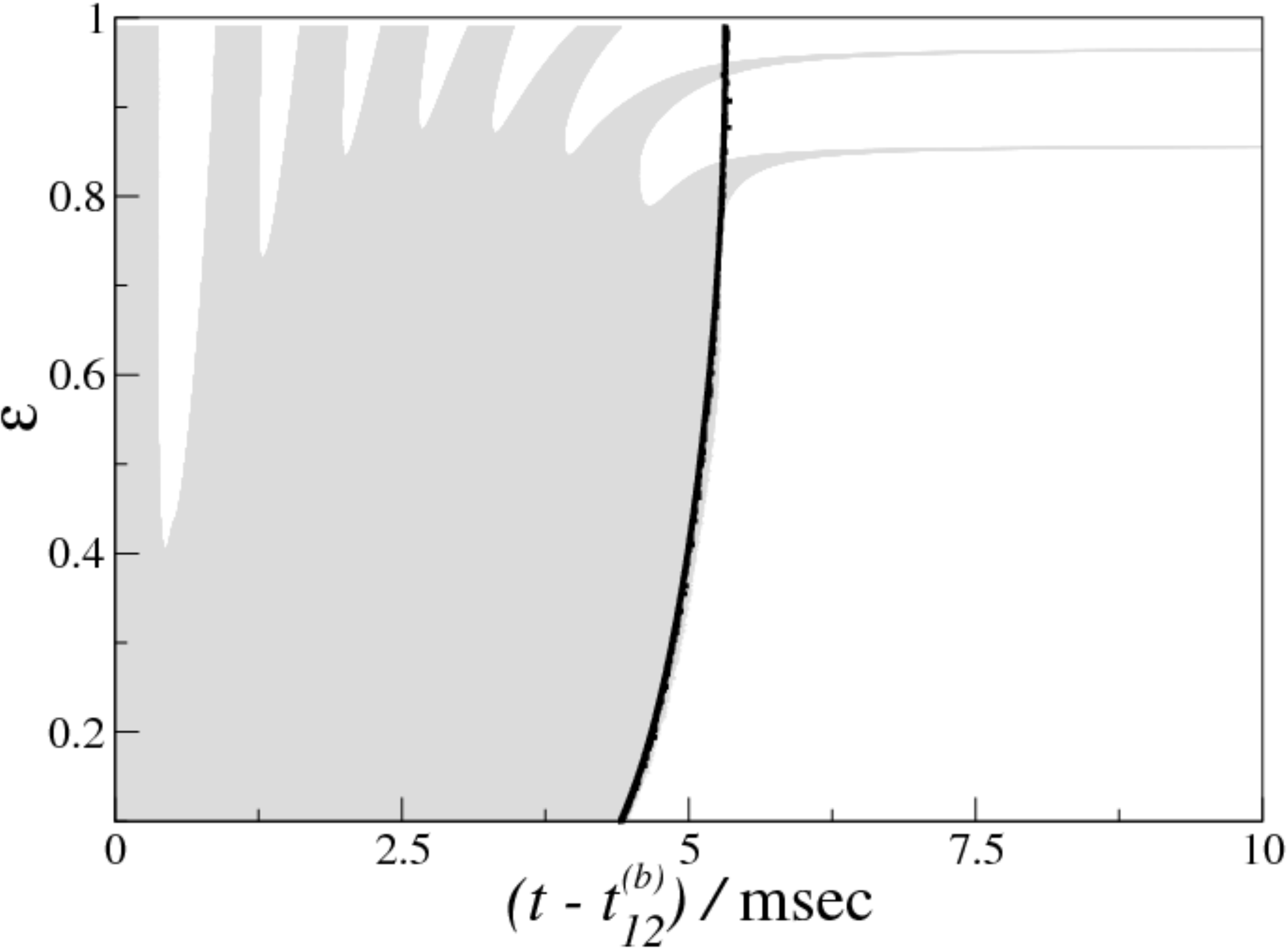}
\caption{Same as Fig. \ref{fig:dash_contacts} but for viscoelastic
spheres. Parameters: ($R_1=10\text{cm}$, $R_2=0.25\text{cm}$, $\Delta
h=0.1\text{mm}$, $Y=5.0\cdot10^7\text{N}/\text{m}^2$, $\nu=0.45$). The black line shows the end of the contact between the lower sphere and the floor for each $\varepsilon$.
}
  \label{fig:visco_contacts}
\end{figure}

Figure \ref{fig:visco_contacts} shows the sequence of collisions, corresponding to Fig. \ref{fig:dash_contacts} for the linear-dashpot force. The figures reveal the same structure of the collision scenario, that is, the viscoelastic force, Eq. \eqref{eq:visc_el_force}, leads to qualitatively the same results as the linear-dashpot model and, thus, 
justifies application of the simplified force Eq. \eqref{eq:dashpot} in Sec. \ref{sec:tbp_lin_dash}.    

\section{Validity of the assumption of a universal coefficient of restitution}
\label{sec:same-material}

For the calculations we assumed that the collisions between the lower sphere and the floor and between the spheres occur via the same coefficient of restitution $\varepsilon$ which allows to consider $\varepsilon$ as a control parameter. This assumption, however, implies also different dissipative constants for the contacts. 

\begin{figure}[t!]
  \centering
  \includegraphics[width=0.99\columnwidth]{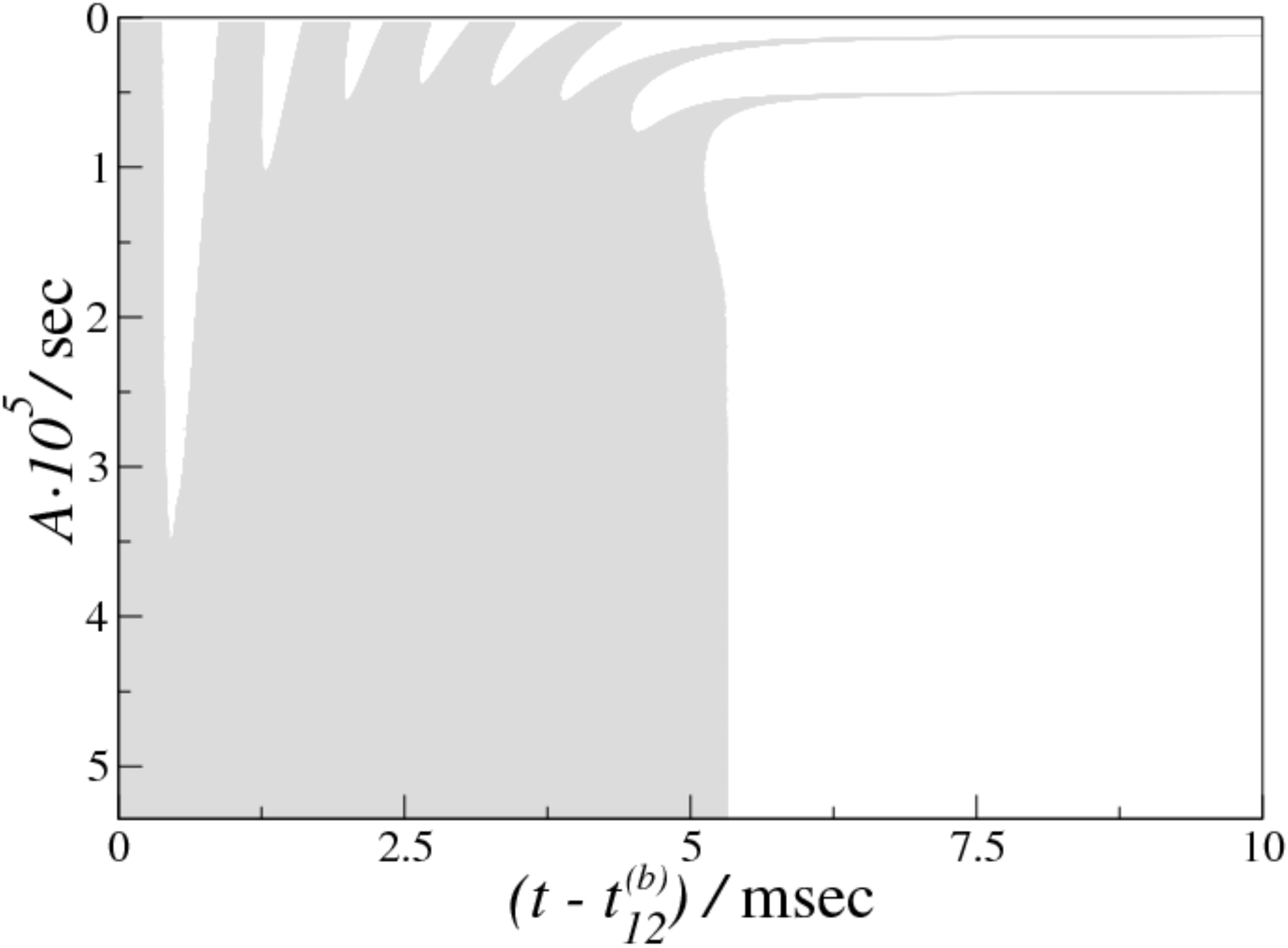}
\caption{Same as Fig. \ref{fig:visco_contacts} but for particles made from the
same material, characterized by the dissipative material parameter $A$.
(Parameters: $R_1=10\text{cm}$, $R_2=0.25\text{cm}$, $\Delta
h=0.1\text{mm}$, $Y=5.0\cdot10^7\text{N}/\text{m}^2$, $\nu=0.45$). 
}
  \label{fig:visco_contacts_au=ao}
\end{figure}

In this Appendix we reproduce Fig. \ref{fig:visco_contacts} once again
with the complementary
assumption of identical material parameters which implies in its turn different coefficients
of restitution $\varepsilon_{01}$ for the particle-floor and $\varepsilon_{12}$
for the particle-particle contact, see Fig. \ref{fig:visco_contacts_au=ao}. 
Thus, we can no longer use $\varepsilon$ as
characteristic value. Instead, to characterize the interaction we use the
dissipative material parameter $A$ which enters the force law, Eq. \eqref{eq:visc_el_force}.

The sequence of collisions shown in Fig. \ref{fig:dash_contacts} has the same structure as for the
assumption of a universal coefficient of restitution with only minor
quantitative differences. Still there occur multiple collisions and consequently
the effects described in Sec. \ref{sec:tbp_lin_dash} persist. Hence the
assumption of a universal coefficient of restitution is justified.


\bibstyle{apsrev}
\bibliography{MULTIPLE}

\end{document}